\documentclass[reprint,twocolumn,showpacs,amsmath,amssymb,superscriptaddress,aps,prl]{revtex4-1}

\usepackage{graphicx}
\usepackage[english]{babel}

\usepackage{amsmath}
\usepackage{amsfonts}
\usepackage{amssymb}
\usepackage{graphicx}
\usepackage{epsfig}
\usepackage{color}
\usepackage{empheq}
\usepackage{epstopdf}
\newcommand{\RR}{\right}
\newcommand{\LL}{\left}

\newcommand{\dg}{\dagger}
\newcommand{\eref}[1]{Eq.~(\ref{#1})}
\newcommand{\fref}[1]{Fig.~\ref{#1}}

\newcommand{\vari}{}

\begin{document}

\title{Dynamically creating tripartite resonance and dark modes in a multimode optomechanical system}
\author{Erno Damsk\"agg}
\address{Department of Applied Physics, Aalto University, 
PO Box 11100, FI-00076 Aalto, Finland}
\author{Juha-Matti Pirkkalainen}
\address{Department of Applied Physics, Aalto University, 
PO Box 11100, FI-00076 Aalto, Finland}
\author{ Mika A. Sillanp\"{a}\"{a}}
\address{Department of Applied Physics, Aalto University, 
PO Box 11100, FI-00076 Aalto, Finland}
\email[]{erno.damskagg@aalto.fi}
\begin{abstract}
We study a multimode  optomechanical system where two mechanical oscillators are coupled to an electromagnetic cavity. Previously it has been shown that if the mechanical resonances have nearly equal frequencies, one can make the oscillators to interact via the cavity by strong pumping with a coherent pump tone. One can view the interaction also as emergence of an electromagnetically dark mode which gets asymptotically decoupled from the cavity and has a linewidth much smaller than that of the bare cavity. The narrow linewidth and long lifetime of the dark mode could be advantageous for example in information storage and processing. Here we investigate the possibility to create dark modes dynamically using two pump tones. We show that if the mechanical frequencies are intrinsically different, one can bring the mechanical oscillators and the cavity on-resonance and thus create a dark mode by double sideband pumping of the cavity. We realize the scheme in a microwave optomechanical device employing two drum oscillators with unmatched frequencies, $\omega_{1} / 2\pi = 8.1 \, \mathrm{MHz}$ and $\omega_{2} / 2\pi = 14.2 \, \mathrm{MHz}$. We also observe a breakdown of the rotating-wave approximation, most pronounced in another device where the mechanical frequencies are close to each other. \end{abstract}

\maketitle

%
\vspace{2pc}


\section{Introduction}
Cavity optomechanical systems couple light or microwave field and mechanical vibrations in a confined volume.  Optomechanical systems have been actively utilized to study quantum effects in macroscopic systems. Freezing of the thermal motion of the mechanical oscillator close to the quantum ground state has been observed \cite{Teufel2011b, AspelmeyerCool11,Painter2015Gnd,Painter2016SiN}, as well as entanglement between the cavity and a mechanical oscillator \cite{LehnertEnta2013}. More recently, microwave optomechanical systems were taken advantage of in squeezing the quantum noise of mechanical motion \cite{SchwabSqueeze,Squeeze,TeufelSqueeze}. A natural extension to studies involving one cavity and one oscillator, is to link more cavities or more mechanical oscillators to participate in interesting dynamics. Various theoretical studies have focused on the creation of complicated collective states \cite{Heidmann2005,Vitali08,Xuereb2012PRL,Marquardt2013Array,Meystre2013Multi,China2013multi,WoolleyBAE,ClerkEnt2014,Vitali2015array}. In experimental work, cavity-mediated coupling of mechanical oscillators has been observed several times in the optical domain \cite{PainterMix2010,Lipson2012sync,Harris2014TwoMode,TwoMode2014,Marin2016TwoSqu}, or with microwaves \cite{multimode2012}. Several studies have focused on several cavity modes coupled to a single mechanical oscillator \cite{Wang2012Dark,Painter2012twocav,Clerk2012Dark,Tian2012interf, Monifi:2016aa}. This scheme can reveal a mechanically "dark" mode, where the energy predominantly resides in the cavities. 

In a system where two mechanical oscillators are coupled to a single cavity, with strong enough coupling, the eigenmodes consist of two modes which are cavity-like, and one mechanical-like mode which is optically dark \cite{multimode2012}. Below we refer the latter as the dark mode. Because in the dark mode the energy is in the long-lived mechanical oscillators, it has much smaller linewidth and larger lifetime as compared to the electromagnetic mode. These properties are attractive for creating narrowband filters for signal processing, or possibly for transferring quantum information to the dark mode. Experimental evidence of the formation of a dark mode has been obtained in a microwave optomechanical experiment \cite{multimode2012}. The problem for carrying out interesting experiments with dark modes is that in order to reach the  case where the energy is mostly in the mechanical oscillations, one needs to have two oscillators with nearly equal eigenfrequencies, in practice differing less than a few \% from each other. Given a typical spread in manufacturing by nanofabrication, it is challenging to obtain such tolerances. In this work we address this issue by demonstrating an approach for the creation of dark modes by dynamically bringing two mechanical oscillators on-resonance by pumping with two microwave pump tones. There is no need to have the intrinsic mechanical frequencies close to each other. Exactly the opposite holds true; the further the frequencies are from each other, the higher fidelity the dark mode has.

\section{Theoretical discussion}

The system we study consists of two mechanical oscillators $j = 1, 2$ that have no direct coupling. They are both coupled to an electromagnetic cavity via the standard radiation-pressure interaction. The cavity can in principle be either in the optical or microwave frequency range, but in the practical part of this work we use the latter (see \fref{fig:meassetup}a). The system is described using the cavity optomechanical Hamiltonian
\begin{equation}
\begin{split}
H = & \omega_{c}a^{\dagger}a +  \sum_{j=1,2}\omega_{j}b_{j}^{\dagger}b_{j} 
- a^\dg a\sum_{j=1,2}g_{0j}x_{0j}(b_{j}^{\dagger} + b_{j}) \\
& + H_{p} \,.
\label{eq:multimodeHam}
\end{split}
\end{equation}
Here, $\omega_{c}$ is the frequency of the cavity, $\omega_{j}$ are the frequencies of the mechanical oscillators, $a^\dg, a$ are the cavity operators, and $b_{j}^{\dagger}, b_{j}$ are the operators for the oscillators. The mechanical zero-point amplitudes are denoted by $x_{0j}$, and $g_{0j} = \frac{\partial  \omega_c}{\partial x_j}$ are the single-photon coupling energies. $H_{p}$ is the pump Hamiltonian which in the most general case considered in this work consists of two different pump tones:
\begin{equation}
\begin{split}
 H_{p} =  d_{1} (a^\dg +a )\cos{\omega_{p1}t} + d_{2} (a^\dg +a)\cos{\omega_{p2}t} \,,\label{eq:pumpham}
\end{split}
\end{equation}
where $d_{j}$ relate to the power of each pump tone. The frequency $\omega_{p1}$ can be thought as being associated to pumping oscillator 1, and similarly for $\omega_{p2}$ and oscillator 2. \vari{As displayed in the frequency scheme  in \fref{fig:meassetup}b, the two pump tones can be detuned from the respective sideband resonance conditions $\omega_c - \omega_j$ ($j = 1,2$) by the amounts $\delta_j$, such that $\omega_{pj} = \omega_{c} - \omega_{j} + \delta_{j}$.}

Let us first review the basic case where only a single pump tone is applied, so for for the moment we take $d_{2} = 0$ in \eref{eq:pumpham}. This scheme has been discussed several times in the literature \cite{PainterMix2010,multimode2012}.  The intense pump tone induces a field amplitude $\alpha$ and the photon number $n_p=|\alpha|^2 \gg 1$. After linearization of \eref{eq:pumpham} in a frame rotating with the pump tone, one obtains the single-pump Hamiltonian
\begin{equation}
\begin{split}
H_1 = & -\Delta a^{\dagger}a +  \sum_{j=1,2}\omega_{j}b_{j}^{\dagger}b_{j} 
- \LL( a^{\dagger} + a \RR) \sum_{j=1,2}G_{j} \LL(b_{j}^{\dagger} + b_{j} \RR)
\label{eq:singlepumpHam}
\end{split}
\end{equation}
In \eref{eq:singlepumpHam},  detuning of the pump tone from the cavity frequency is defined as $\Delta = \omega_{p1} - \omega_c$, and the effective coupling is $G_j = g_{0j} \sqrt{n_p}$. The effective frequency $-\Delta$ of the cavity is tunable by the pump tone frequency. The eigenfrequencies of the system, and the corresponding damping rates can be obtained from the equations of motion corresponding to \eref{eq:singlepumpHam}. An interesting case is that with equal mechanical oscillator frequencies, $\omega_{1} = \omega_{2}$. Then \eref{eq:singlepumpHam} describes three oscillators which become co-resonant when the pump tone is applied at the red-sideband frequency  $\Delta = - \omega_{1}= - \omega_{2} $. In this case, one can show that the relative motion of the mechanical oscillators weighed by the couplings, namely $x_d = \LL( G_2 x_1 - G_1 x_2 \RR)/\LL( G_2 +  G_1 \RR)$ is an eigenmode that is uncoupled from the cavity. It is the dark mode which has the frequency $\omega_d = \LL( \omega_1 + \omega_2\RR)/2$ and damping rate $\gamma_d = \LL( \gamma_1 + \gamma_2 \RR)/2$. If the mechanical frequencies are not equal, the mode structure is more complicated, and the dark mode shows up properly only in the strong-coupling limit $G_j \gg \kappa$. A representative plot of the damping rates of the eigenmodes, involving some detuning of the oscillators, is shown in \fref{fig:meassetup}c.

In order to create a true tripartite resonance, in practice one needs to be able to tune the mechanical frequencies in situ. This is possible by voltage gating, but adds complexity to the design. In the following we focus on the two-pump driving conditions which allow, as we will show, dynamic tuning of the effective mechanical frequencies in the model. In \eref{eq:pumpham} we now include both pump tones, $d_1, d_2 \neq 0$. The field inside the cavity is the sum of the pump-induced, large fields $\alpha_{j}$, plus small fluctuations: $\vari{a  = \alpha_{1} e^{i\omega_{p1}t} + \alpha_{2} e^{i\omega_{p2}t}  + \delta a}$. We shift  \eref{eq:multimodeHam} to a rotating frame with respect to $\omega_{p1}\delta a^{\dagger} \delta a + \Delta_{p} b_{2}^{\dagger}b_{2}$, where $\Delta_{p} = \omega_{p1} - \omega_{p2}$ is the difference between the pump tone frequencies. \vari{We carry out the transformation, linearization and rotating-wave approximation (RWA). For the sake of simple notation, we also denote in the following $\delta a \Rightarrow a$. We get the Hamiltonian}
\begin{equation}
\begin{split}
H = &(\omega_1 - \delta_1) a^{\dagger}a +  \omega_{1}b_{1}^{\dagger}b_{1} +  (\omega_{2} - \Delta_{p})b_{2}^{\dagger}b_{2}+ \\ 
& + G_1 \LL( a^\dg b_1 + a b_1^{\dg}  \RR)+ G_2 \LL(a^\dg b_2 + a b_2^{\dg}\RR) \,.
\label{eq:twopumpHam}
\end{split}
\end{equation}
Apart from neglected Stokes processes $\propto a b_j $+ h.c., \eref{eq:twopumpHam} is of the same form as \eref{eq:singlepumpHam}. The frequency of oscillator 2 is now tunable by the frequency of pump tone 2.
We can individually put any number of the subsystems on resonance by changing the frequencies of the two pump tones. In particular, one can create a dark mode by playing with the pump parameters. 

The crucial assumption in obtaining the time-independent \eref{eq:twopumpHam} is the RWA. We have neglected sidebands at the frequency $\omega_{p2} - \omega_{p1} \simeq \omega_{2} - \omega_{1}$. \vari{Because the neglected sidebands have  equally strong prefactors as the retained coupling terms in \eref{eq:twopumpHam}, this assumption is justified if the neglected sidebands are strongly suppressed by the cavity response, requiring that $\omega_{2} - \omega_{1} \gg \kappa$. Thus, if the mechanical resonators have eigenfrequencies close to each other, the time-independence of the obtained Hamiltonian \eref{eq:twopumpHam} is not fulfilled.} Physically, the most important assumption is that each pump tone couples only to the respective oscillator. Ignoring the Stokes processes means that \eref{eq:twopumpHam} holds only when both the pump tones are on the red-detuned side.

Obtaining a transmission or reflection spectrum from \eref{eq:twopumpHam} is a nontrivial but standard exercise in the input-output formalism. Here we refer to the supplementary of Ref.~\cite{MechAmpPaper}. We note that the measured transmission coefficient does not necessarily directly indicate where a given eigenmode is. However, we argue that in the absence of sharp features other than the dark mode in the spectrum that might contribute interference, the microwave response provides a good reconstruction of the dark mode.

\section{Experimental results}
\begin{table*}
\caption{Characteristics of devices measured in this work. Shown are the cavity frequencies $\omega_{c} $,  mechanical oscillator frequencies $\omega_{j}$, cavity decay rates $\kappa$, mechanical decay rates $\gamma_{j}$, and the ratio of single photon coupling rates $g_{1} / g_{2}$.}
\begin{tabular}{|l|l|l|l|l|l|l|l|l|l|l|}
\hline
Device & $\omega_{c} / 2\pi$& $\omega_{1}  / 2\pi$ & $\omega_{2}  / 2\pi$ & $\kappa/2\pi$ & $\kappa_{I}/2\pi$ & $\kappa_{Ei} / 2\pi$& $\kappa_{Eo} / 2\pi$ &$\gamma_{1} $& $\gamma_{2}$& $g_{1} / g_{2}$\\
& (GHz)& (MHz) & (MHz) & (kHz) & (kHz) & (kHz) & (kHz) & (Hz) & (Hz) & \\
\hline
1 &6.9635 & 8.11 &14.17 & 620 & 130 & 180 & 310 & 64 & 110 & 2.2 \\
\hline
2 & 4.9979 & 12.93& 14.48& 815 & 85 & 10 & 720  &1196 & 1308 & 0.65  \\
\hline
\end{tabular}
\end{table*}

\begin{figure*}
\includegraphics[height=10cm]{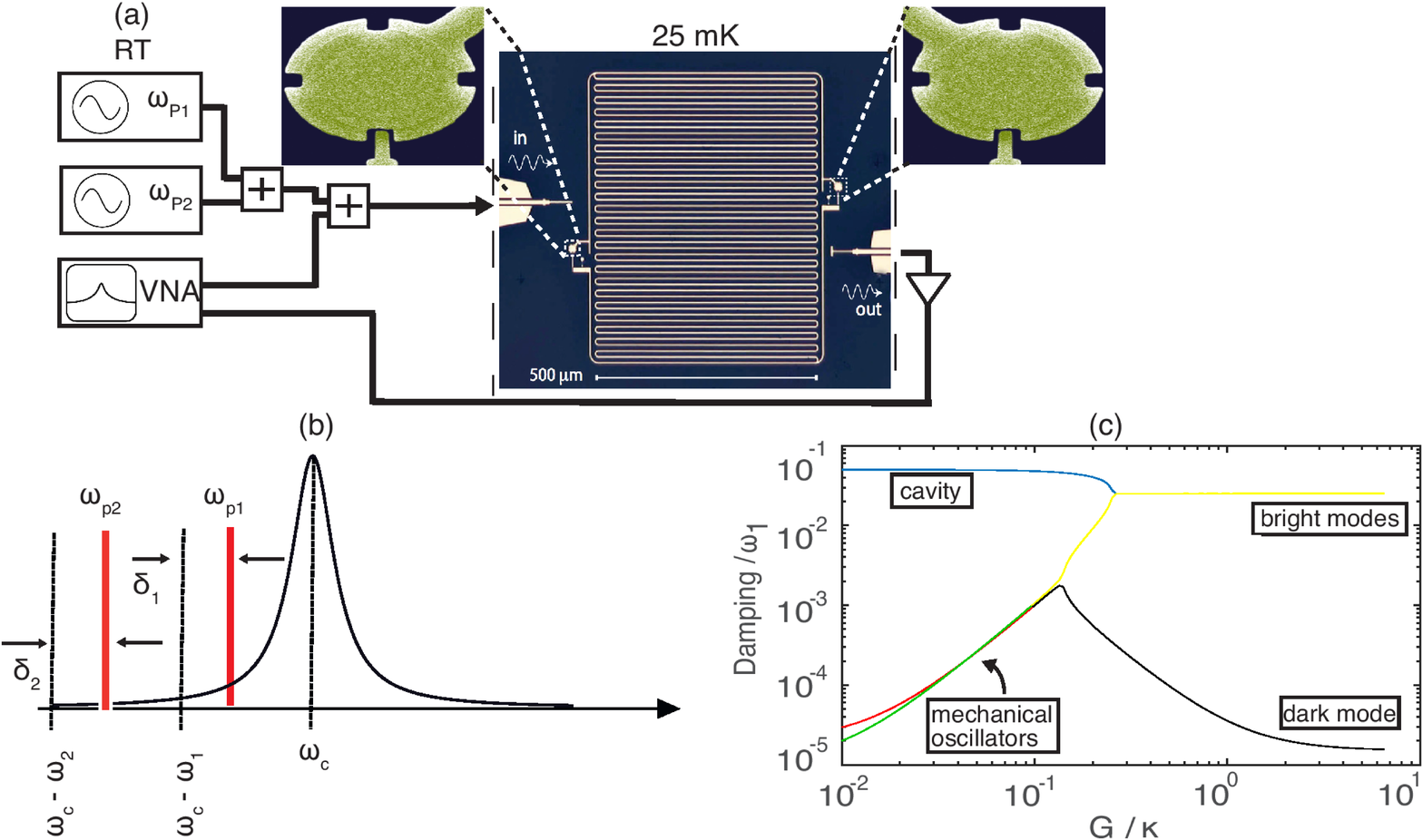}
\caption{(a) Schematics of the experiment. In the sample there are two drum oscillators which in the main micrograph are marked with small dashed boxes on opposite sides of the aluminum transmission line resonator. The enlarged views depict a drum oscillator. Room temperature measurement setup is shown on the left. Shown are two signal generators which are used as source of the microwave pump tones, and a vector network analyzer (VNA). (b) Frequency scheme used in the measurements. Cavity is pumped with two pump tones of  frequencies $\omega_{p1,2}$ which are set near-resonant at the red-sideband frequencies. (c) Calculated damping rate of eigenmodes of the tripartite system as a function of pump power. The parameters for the plot are $\omega_1/\omega_2 = 0.998$, $\gamma_1/\omega_1 = 10^{-5}$, $\gamma_2/\omega_1 = 2 \cdot 10^{-5}$, $\omega_{1}/\kappa = 20$.}
\label{fig:meassetup}
\end{figure*}

In the experiments we use devices where the mechanical oscillators are two 120 nm thick aluminum drum membranes (shown in Fig.\ref{fig:meassetup}a), which are capacitive coupled to an on-chip superconducting microwave resonator that acts as a cavity. We discuss two different samples, numbered 1 and 2, which  differ in the frequency spacing of the oscillators, leading to qualitatively different physics. Experiments were conducted in a dilution cryostat at a temperature 25 mK.  As depicted in \fref{fig:meassetup}a, the room temperature measurement setup consisted of two signal generators and a vector network analyzer, which is used to record the microwave transmission parameter $S_{21}$. The two signal generators acted as the  sources for the two independent pump tones. 

The relevant device parameters are listed in Table 1. The losses of the cavity are classified in three different origins. The internal losses are labeled with the symbol $\kappa_{I}$, and the cavity is coupled to transmission lines through input and output ports with decay rates $\kappa_{Ei}$ and $\kappa_{Eo}$, respectively. The losses sum up to the total cavity decay rate $\kappa = \kappa_{I} + \kappa_{Ei}+ \kappa_{Eo}$.

\begin{figure}
       \includegraphics[height=9cm]{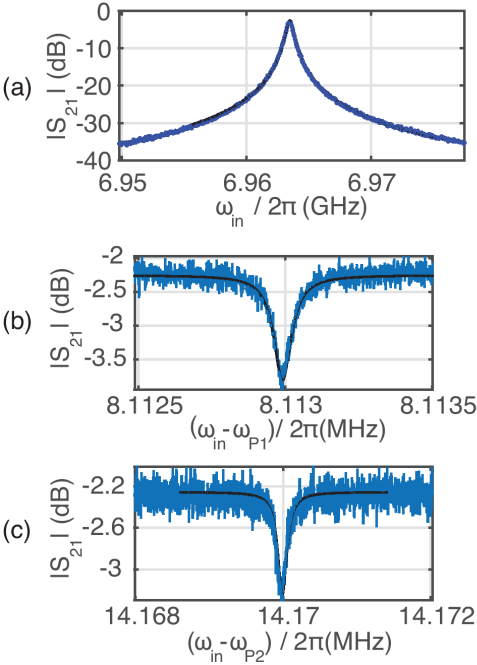}
    \caption{Device 1, basic characterization: (a) Cavity transmission spectrum with no pump tones applied. (b) and (c), Resonances of  mechanical oscillators 1 and 2, respectively, when either of them is weakly pumped at the red-sideband. Transmission spectrum is fitted (black curves) to each plot.}
    \label{fig:devicefits}
\end{figure}

We first discuss device 1. As we will show, this device exhibits distinct dark modes in its spectrum and is well described by \eref{eq:twopumpHam}. The basic transmission measurement involving only the probe tone reveals the cavity resonance peak (\fref{fig:devicefits}a), and narrow dips corresponding to the mechanical oscillators when a single weak pump tone is applied (\fref{fig:devicefits}b,c). In the latter measurement, we adjust the pump power such that the resonance dips are roughly equally tall. This requires about 10 dB more generator power applied to oscillator 2. Considering also the effect of a larger detuning from cavity for oscillator 2, we obtain the ratio of bare couplings $g_1/g_2 \simeq 2.2$. In what follows, we use this information to balance roughly equal effective couplings which is the most reasonable choice for constructing the physics.

\begin{figure}
\includegraphics[height=7cm]{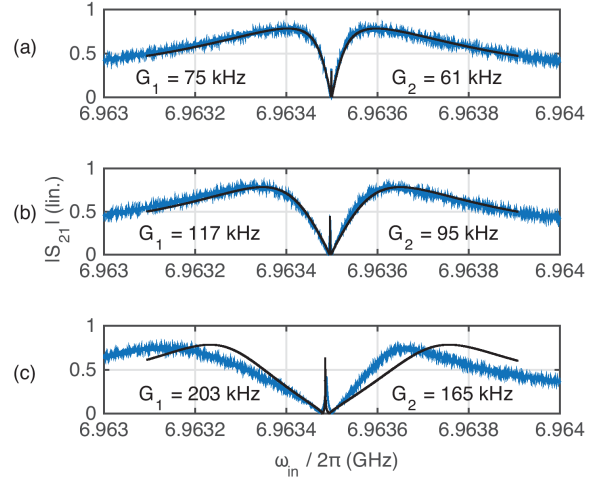}
\caption{Device 1: Both mechanical oscillators pumped at sideband-resonance $\delta_1=\delta_2=0$. The pump power is increased from (a) to (c), while keeping the ratio $G_1/G_2$ constant. Theory curves (black) are fitted to the measured spectrum.}
\label{fig:measplots}
\end{figure}

We next apply both pump tones such that both $d_1$ and $d_2$ in \eref{eq:pumpham} are non-zero. We first apply the pump tones nominally at the sideband resonance  of either oscillator, namely \vari{$\delta_1=\delta_2 = 0$}, thus creating the tripartite on-resonance system. In \fref{fig:measplots}, we display data which, first of all, shows splitting of the cavity peak suggesting onset of the strong-coupling regime. In the center dip, one can identify a sharp peak which we associate to the dark mode. The agreement to the theoretical expectation based on \eref{eq:twopumpHam} in the overall profile is good except at the highest power in \fref{fig:measplots}c, where the response becomes asymmetric because of the cavity frequency develops power-dependence, hence detuning both pump tones from the sideband resonance. The widths of the center peaks corresponding to the dark mode are extracted as $1.2$ kHz, $1.8$ kHz and  $4.9$ kHz, in \fref{fig:measplots}a-c, respectively. These values are clearly higher than the anticipated $\gamma/2\pi \lesssim 100$ Hz independent on the pump power. We attribute the difference to a breakdown of the RWA towards high pump powers. Moreover, in the ideal model, the dark mode is truly dark and hence does not leave a trace whatsoever in a measurement though the cavity. We will return to the latter point below.

\begin{figure}
\includegraphics[height=14cm]{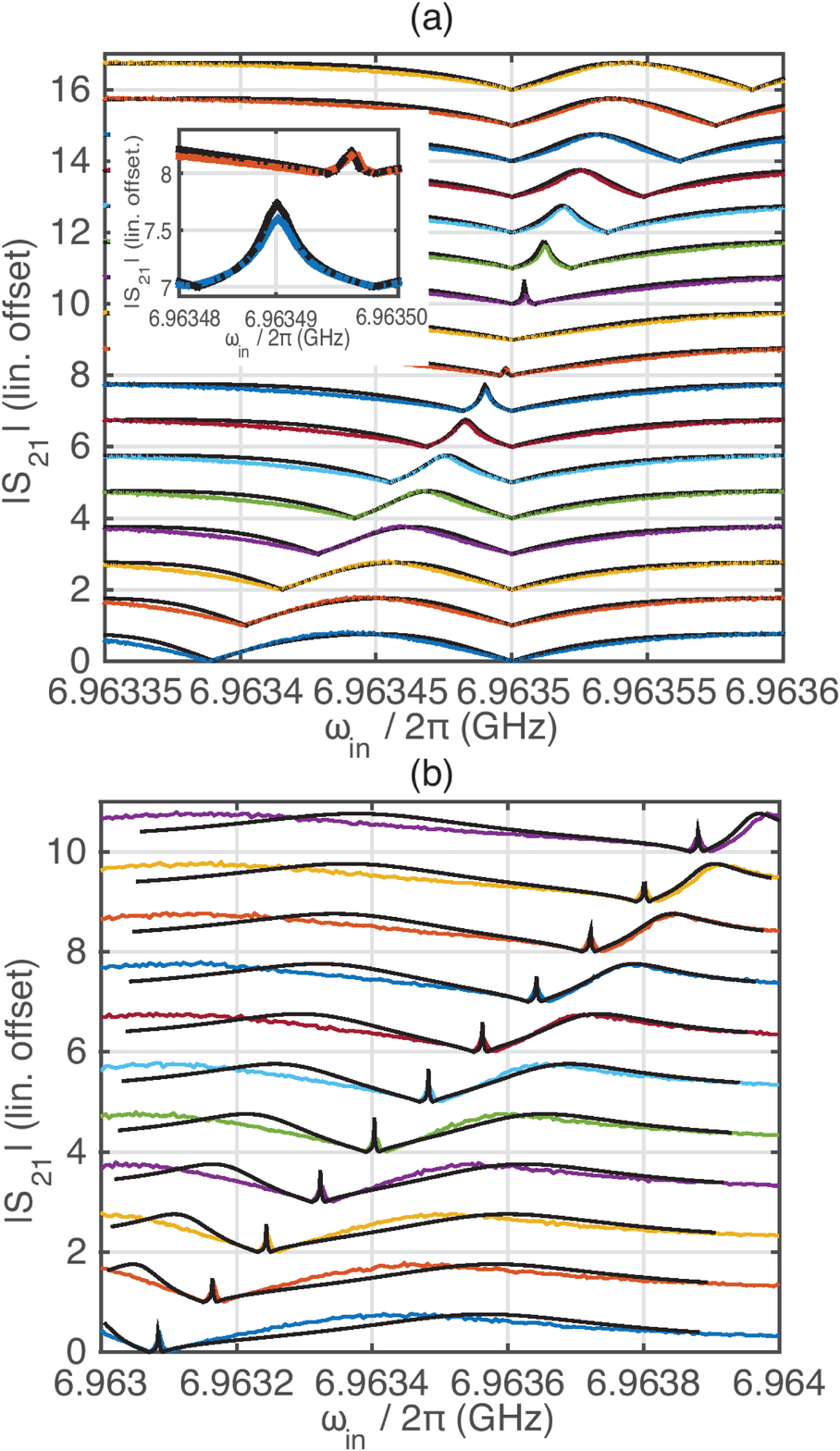}
\caption{Device 1: Fixed pump powers, varying detunings. (a) Oscillator 1 was pumped with  $G_{1}/2\pi \simeq  8.5 \cdot 10^{4}$ kHz, and oscillator 2 at $G_{2}/2\pi \simeq   7.7 \cdot 10^{4}$ kHz. The pump tone frequency $\omega_{p2}$ was kept at sideband resonance, $\delta_{{2}} = 0$. The frequency of pump tone 1 was stepped corresponding to $\delta_{{1}} = -100$ kHz ... 100 kHz, from bottom to top. (b)  Both pump tone frequencies stepped with the same spacing, $\delta_{1} = \delta_{2} \simeq $ - 400 kHz ... 400 kHz, from bottom to top. The pump amplitudes are $G_{1}/2\pi \simeq  170$ kHz, and  $G_{2}/2\pi \simeq 130$ kHz. Theory prediction (black) is fitted to the spectra.}
\label{fig:sweeps}
\end{figure}

We now show how to adjust the system on or off resonance by the frequency of pump tone 2. As seen in \eref{eq:twopumpHam}, the effective mode frequency $\omega_{2} - \Delta_{p}$ of the mechanical oscillator 2 can be controlled by either of the pump tones. In the situation corresponding to \fref{fig:sweeps}a, we keep pump powers constant. The frequency of pump tone 2 is fixed at the sideband-resonance detuning \vari{$\delta_2 = 0$}, whereas  pump tone 1 is swept in frequency around the \vari{sideband-resonance condition $\delta_1 \sim 0$}. At the largest detunings shown, the two oscillators show up as two roughly independent dips in the spectrum. When the detuning is swept through the co-resonance point (middle curves), the sharp peak reflecting the dark mode clearly appears. Another operation which is possible by tuning the pump tone frequencies is to change both frequencies such that their difference $\Delta_{p}$ stays constant. Based on \eref{eq:twopumpHam}, this amounts to keeping both oscillators resonant with each other, while sweeping the cavity through them. The measured spectrum is shown in \fref{fig:sweeps}b. As predicted, the dark mode position shifts with the pump tone detuning.

\begin{figure}
\includegraphics[height=4cm]{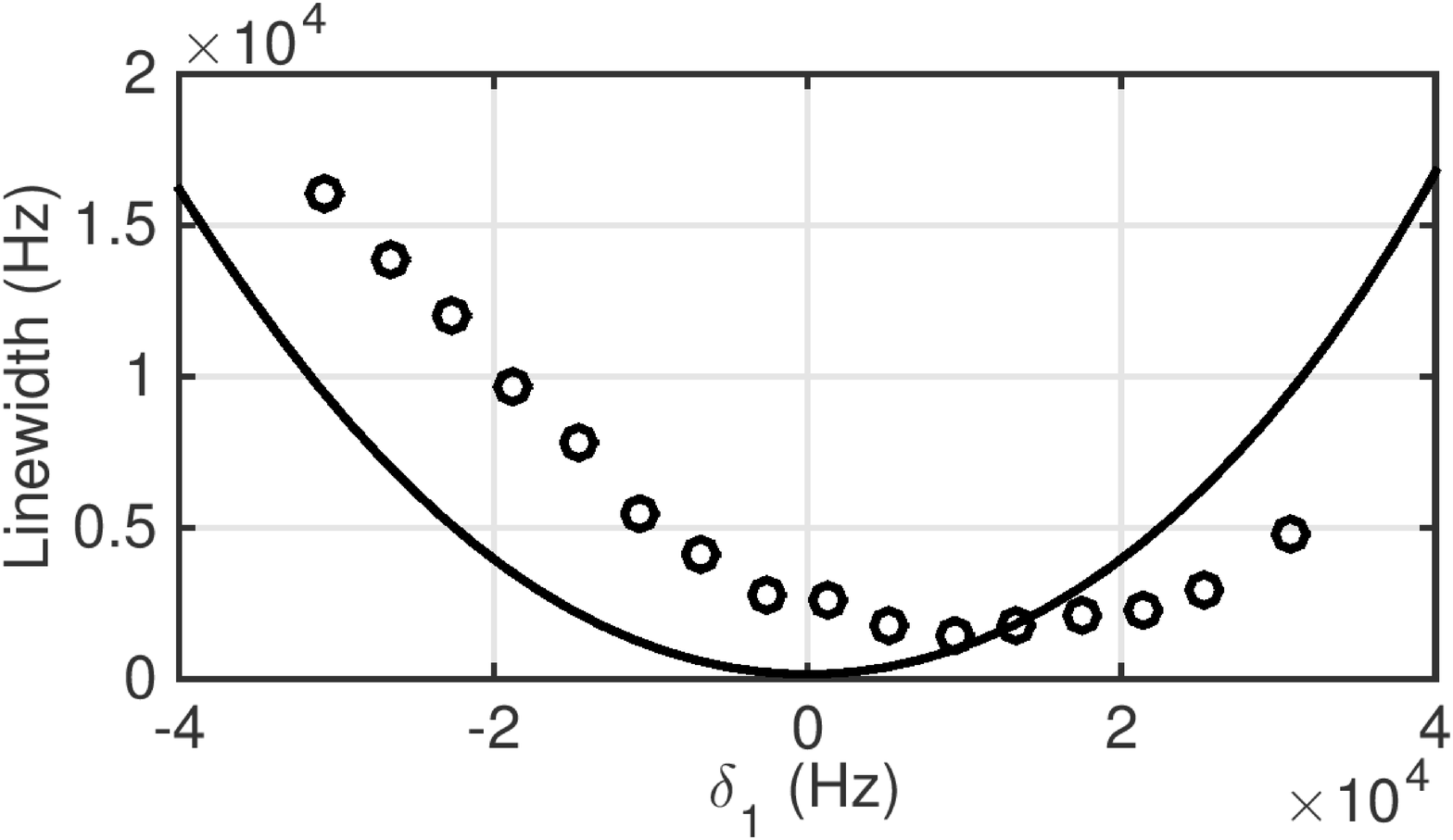}
    \caption{Device 1: Linewidth of the dark mode peak obtained from \fref{fig:sweeps}a. The solid line is a theoretical prediction.}
\label{fig:linewidth}
\end{figure}

We can extract the dark mode peak linewidths from \fref{fig:sweeps}a, and compare to the theoretical predictions following from the equations of motion. This is illustrated in \fref{fig:linewidth}. Apart from not reaching the lowest values of linewidth as predicted by the ideal model, the minimum of the data is shifted from the theoretical pattern. We argue these features are a result of breakdown of the rotating-wave approximation. When approaching the proper dark mode conditions where the energy is not supposed to be in the cavity, the neglected sidebands can become important because their energy is contained in the cavity. The mentioned shift of the co-resonance condition is also within the data shown in \fref{fig:measplots}. Here, the nominal co-resonance was selected in experiment. The corresponding theory curves are taken with parameters shifted by the mentioned minimum shift to allow for a reasonable comparison.

\begin{figure}
\includegraphics[height=10cm]{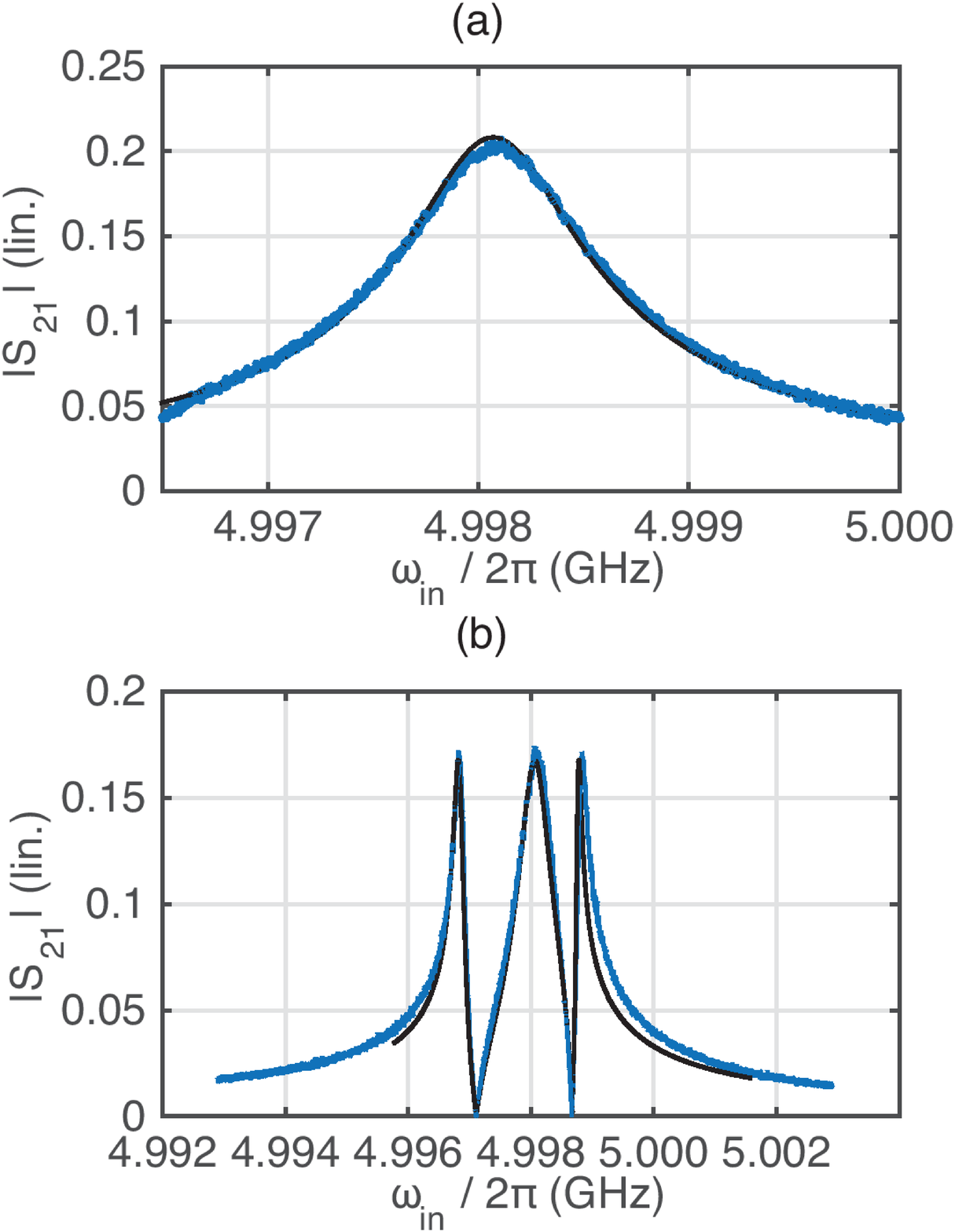}
    \caption{Device 2: (a) Cavity response without pump tones.  (b) Experiment using a single strong pump tone at the frequency $\omega_{p} = \omega_{c} - \frac{\omega_{1} + \omega_{2}}{2}$, with $G$ =  285 kHz. Theory curves (black) are fitted to the response.}
    \label{fig:uusinayte1}
\end{figure}

\begin{figure}
\includegraphics[height=12cm]{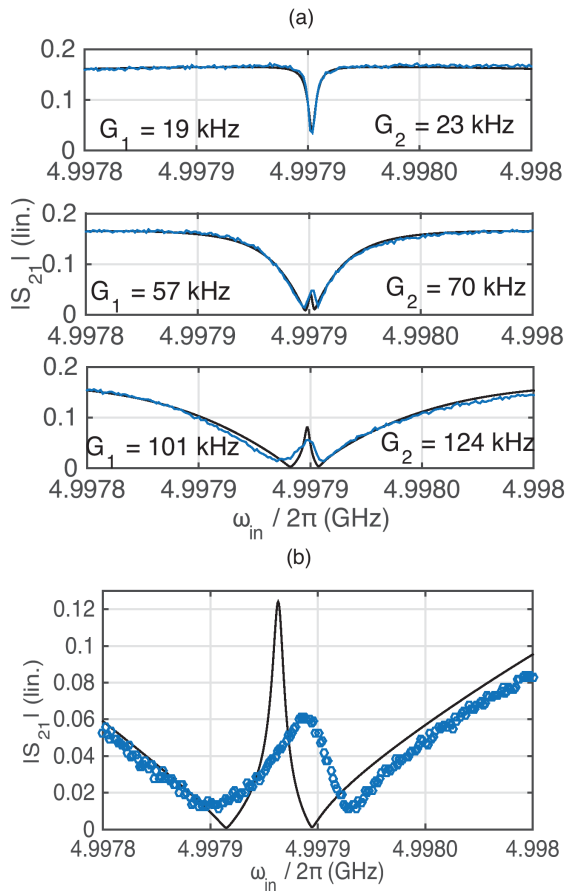}
    \caption{Device 2: (a) Both mechanical oscillators pumped on sideband resonance. The difference between two pump powers is kept constant in each panel. (b) Zoom-in to the dark mode peak.}
    \label{fig:uusinayte2}
\end{figure}

We also investigated another device in which the two mechanical oscillators have resonance frequencies closer to each other, namely $\omega_{1}  / 2\pi = 12.93$ MHz and $\omega_{2}  / 2\pi = 14.48$ MHz. The basic characterization is shown in \fref{fig:uusinayte1}. The detuning of the oscillators is too large as compared to the achievable effective coupling, so that using a single pump tone, a dark mode having an appreciable mechanical component cannot be created. The response shows a broad peak in the center (\fref{fig:uusinayte1}b), which would evolve towards the dark mode at higher pump power. In the double-pump measurement displayed in \fref{fig:uusinayte2}, both pump tones are applied at the sideband resonance  (similar to \fref{fig:measplots} for device 1). There is a qualitative resemblance to \fref{fig:measplots}, however, the center peak is broader. In this device, the dark mode peak is less well described by the theory, because the \vari{time-independence condition of the Hamiltonian in \eref{eq:twopumpHam}} $\omega_{2} - \omega_{1} \gg \kappa$ is not properly satisfied.


\section{Summary}

We have shown how a multimode cavity optomechanical system that is pumped with several coherent pump tones exhibits intriguing collective dynamics. If there are several mechanical oscillation modes coupled to the cavity, two of them can be simultaneously brought on-resonance with the cavity by tailoring the pump tone parameters. The tripartite co-resonance conditions entails a strong interaction between the oscillators which have no direct coupling and which can be physically far separated. All this is possible in principle with an arbitrarily small pump power. The dynamic control is more efficient than what demonstrated previously using a single pump tone. However, care has to be taken to ensure the rotating-wave approximation holds well enough. This requires the intrinsic frequencies to be substantially different as compared to the cavity linewidth. On the other hand, the possibility to bring very unequal oscillators effectively on-resonance could be the most appealing consequence, because oscillators with unmatched frequencies are difficult to connect otherwise. We anticipate the scheme can be extended to dynamical tuning of frequencies in more complex systems which contain more than two oscillators.

\begin{acknowledgments}
We  thank Francesco Massel and Tero Heikkil\"a for useful discussions. This work was supported by the Academy of Finland (CoE LTQ, 275245) and by the European Research Council (615755-CAVITYQPD). The work benefited from the facilities at the Micronova Nanofabrication Center and at the Low Temperature Laboratory infrastructure.
\end{acknowledgments}

\section{References}
\bibliographystyle{iopart-num}

\bibliography{mikabib}

\providecommand{\newblock}{}
\begin{thebibliography}{10}
\expandafter\ifx\csname url\endcsname\relax
  \def\url#1{{\tt #1}}\fi
\expandafter\ifx\csname urlprefix\endcsname\relax\def\urlprefix{URL }\fi
\providecommand{\eprint}[2][]{\url{#2}}

\bibitem{Teufel2011b}
Teufel J~D, Donner T, Li D, Harlow J~W, Allman M~S, Cicak K, Sirois A~J,
  Whittaker J~D, Lehnert K~W and Simmonds R~W 2011 {\em Nature\/} {\bf 475}
  359--363

\bibitem{AspelmeyerCool11}
Chan J, Alegre T~P~M, Safavi-Naeini A~H, Hill J~T, Krause A, Gr\"oblacher S,
  Aspelmeyer M and Painter O 2011 {\em Nature\/} {\bf 478} 89--92

\bibitem{Painter2015Gnd}
Meenehan S~M, Cohen J~D, MacCabe G~S, Marsili F, Shaw M~D and Painter O 2015
  {\em Phys. Rev. X\/} {\bf 5} 041002

\bibitem{Painter2016SiN}
Fink J~M, Kalaee M, Pitanti A, Norte R, Heinzle L, Davanco M, Srinivasan K and
  Painter O 2015 {\em arXiv:1512.04660\/}

\bibitem{LehnertEnta2013}
Palomaki T~A, Teufel J~D, Simmonds R~W and Lehnert K~W 2013 {\em Science\/}
  {\bf 342} 710--713

\bibitem{SchwabSqueeze}
Wollman E~E, Lei C~U, Weinstein A~J, Suh J, Kronwald A, Marquardt F, Clerk A~A
  and Schwab K~C 2015 {\em Science\/} {\bf 349} 952--955

\bibitem{Squeeze}
Pirkkalainen J~M, Damsk\"agg E, Brandt M, Massel F and Sillanp\"a\"a M~A 2015
  {\em Phys. Rev. Lett.\/} {\bf 115} 243601

\bibitem{TeufelSqueeze}
Lecocq F, Clark J~B, Simmonds R~W, Aumentado J and Teufel J~D 2015 {\em Phys.
  Rev. X\/} {\bf 5} 041037

\bibitem{Heidmann2005}
Pinard M, Dantan A, Vitali D, Arcizet O, Briant T and Heidmann A 2005 {\em
  Europhysics Letters\/} {\bf 72} 747

\bibitem{Vitali08}
Genes C, Vitali D and Tombesi P 2008 {\em New Journal of Physics\/} {\bf 10}
  095009

\bibitem{Xuereb2012PRL}
Xuereb A, Genes C and Dantan A 2012 {\em Phys. Rev. Lett.\/} {\bf 109} 223601

\bibitem{Marquardt2013Array}
Ludwig M and Marquardt F 2013 {\em Phys. Rev. Lett.\/} {\bf 111} 073603

\bibitem{Meystre2013Multi}
Seok H, Buchmann L~F, Wright E~M and Meystre P 2013 {\em Phys. Rev. A\/} {\bf
  88} 063850

\bibitem{China2013multi}
Xu X~W, Zhao Y~J and Liu Y~x 2013 {\em Phys. Rev. A\/} {\bf 88} 022325

\bibitem{WoolleyBAE}
Woolley M~J and Clerk A~A 2013 {\em Phys. Rev. A\/} {\bf 87} 063846

\bibitem{ClerkEnt2014}
Woolley M~J and Clerk A~A 2014 {\em Phys. Rev. A\/} {\bf 89} 063805

\bibitem{Vitali2015array}
Zippilli S, Li J and Vitali D 2015 {\em Phys. Rev. A\/} {\bf 92} 032319

\bibitem{PainterMix2010}
Lin Q, Rosenberg J, Chang D, Camacho R, Eichenfield M, Vahala K~J and Painter O
  2010 {\em Nature Photonics\/} {\bf 4} 236--242

\bibitem{Lipson2012sync}
Zhang M, Wiederhecker G~S, Manipatruni S, Barnard A, McEuen P and Lipson M 2012
  {\em Phys. Rev. Lett.\/} {\bf 109} 233906

\bibitem{Harris2014TwoMode}
Shkarin A~B, Flowers-Jacobs N~E, Hoch S~W, Kashkanova A~D, Deutsch C, Reichel J
  and Harris J~G~E 2014 {\em Phys. Rev. Lett.\/} {\bf 112} 013602

\bibitem{TwoMode2014}
Fu H, Mao T~H, Li Y, Ding J~F, Li J~D and Cao G 2014 {\em Appl. Phys. Lett.\/}
  {\bf 105} 014108

\bibitem{Marin2016TwoSqu}
Pontin A, Bonaldi M, Borrielli A, Marconi L, Marino F, Pandraud G, Prodi G~A,
  Sarro P~M, Serra E and Marin F 2016 {\em Phys. Rev. Lett.\/} {\bf 116} 103601

\bibitem{multimode2012}
Massel F, Cho S~U, Pirkkalainen J~M, Hakonen P~J, Heikkil{\"a} T~T and
  Sillanp{\"a}{\"a} M~A 2012 {\em Nat. Commun.\/} {\bf 3} 987

\bibitem{Wang2012Dark}
Dong C, Fiore V, Kuzyk M~C and Wang H 2012 {\em Science\/} {\bf 338} 1609--1613

\bibitem{Painter2012twocav}
Hill J~T, Safavi-Naeini A~H, Chan J and Painter O 2012 {\em Nat. Commun.\/}
  {\bf 3} 1196

\bibitem{Clerk2012Dark}
Wang Y~D and Clerk A~A 2012 {\em New Journal of Physics\/} {\bf 14} 105010

\bibitem{Tian2012interf}
Tian L 2012 {\em Phys. Rev. Lett.\/} {\bf 108} 153604

\bibitem{Monifi:2016aa}
Monifi F, Zhang J, Kaya {\"O}, Peng B, Liu Y~x, Bo F, Nori F and Yang L 2016
  {\em Nat Photon\/} {\bf 10} 399--405

\bibitem{MechAmpPaper}
Massel F, Heikkil{\"a} T~T, Pirkkalainen J~M, Cho S~U, Saloniemi H, Hakonen P~J
  and Sillanp{\"a}{\"a} M~A 2011 {\em Nature\/} {\bf 480} 351--354

\end{thebibliography}

\end{document}